\documentclass[12pt, letterpaper]{article}
\usepackage[utf8]{inputenc}
\usepackage{authblk, sublabel}
\usepackage{amsmath, amssymb}
\usepackage{graphicx}
\usepackage{pstricks}
\usepackage[margin=0.75in]{geometry}
\usepackage{color}
\usepackage[normalem]{ulem}

\renewcommand{\Im}{\mathrm{Im}}
\renewcommand{\Re}{\mathrm{Re}}
\newcommand{\dfourx}{\mathrm{d}^4 x}
\newcommand{\dt}{\mathrm{d}t}
\newcommand{\dqsq}{\mathrm{d}q^2}
\newcommand{\lsr}{\mathcal{R}}
\newcommand{\fesr}{\mathcal{F}}
\newcommand{\gsr}{\mathcal{G}}
\newcommand{\vev}[1]{\ensuremath{\langle #1\rangle}}
\newcommand{\aGG}{\vev{\alpha G^2}}
\newcommand{\gggGGG}{\vev{g^3 G^3}}
\newcommand{\gev}{\ensuremath{\text{GeV}}}
\newcommand{\ie}{\textit{i.e.}}
\newcommand{\eg}{\textit{e.g.}}

\newcommand{\add}[1]{#1}

\title{Numerically computing QCD Laplace sum-rules using pySecDec}
\author[1]{Steven Esau\footnote{skesau@uwaterloo.ca}}
\author[1]{Derek Harnett\footnote{derek.harnett@ufv.ca}}
\affil[1]{Department of Physics\\ University of the Fraser Valley\\ Abbotsford, BC, V2S 7M8, Canada
}

\begin{document}
\maketitle
\begin{abstract}
pySecDec is a program that numerically calculates dimensionally regularized integrals. 
We use pySecDec to compute QCD Laplace sum-rules for pseudoscalar (\ie, $J^{PC}=0^{-+}$) 
charmonium hybrids, and compare the results to sum-rules computed using analytic results
for dimensionally regularized integrals.
We find that the errors due to
the use of numerical integration methods 
are negligible compared to the uncertainties in the sum-rules 
stemming from the uncertainties in the parameters of QCD, 
\eg, the coupling constant, quark masses, and condensate values.
\add{Also, we demonstrate that numerical integration 
methods can be used 
to calculate finite-energy and Gaussian sum-rules in addition 
to Laplace sum-rules.}
\end{abstract}

\section{Introduction}
QCD sum-rules
are a well-established technique for predicting hadron properties from the 
QCD Lagrangian~\cite{Shifman:1978by,Shifman:1978bx,Reinders:1984sr,Narison:2007spatmp}.
QCD sum-rules are transformed dispersion relations that relate a correlation function 
of two local currents to the correlation function's own imaginary part.
The correlator is computed within the operator product expansion (OPE)
in which perturbation theory is supplemented by nonperturbative corrections, each
correction being a product of a perturbatively computed Wilson coefficient
and a nonzero vacuum expectation value of a local operator, 
\ie, a condensate~\cite{Wilson:1969zs}.
The imaginary part of the correlator, the hadronic spectral function, is 
expressed in terms of hadron properties and a continuum threshold.
Hence, QCD sum-rules relate parameters of QCD (\eg, the strong coupling, quark
masses, and condensates) to parameters of hadronic physics (\eg, masses, widths, 
and hadronic couplings)
in a quantitative expression of quark-hadron duality.

When computing Wilson coefficients (including perturbation theory), 
we usually encounter divergent loop integrals.
Such integrals are often handled using dimensional regularization.
If the integrals come from Feynman diagrams with 
small numbers of loops, external lines, and/or distinct masses, then analytic expressions
for them can often be calculated or found in the literature 
(\eg,~\cite{PascualTarrach1984,BoosDavydychev1991,Davydychev:1990cq,BroadhurstFleischerTarasov1993}).
But, as the complexity of Feynman diagrams increases, computing closed-form
analytic expressions for the needed dimensionally regularized integrals can become 
prohibitively difficult, effectively limiting the applicability of the QCD sum-rules
methodology.

pySecDec is an open source Python and C++ program 
(which makes use of FORM~\cite{Vermaseren:2000nd,Kuipers:2013pba,Ruijl:2017dtg}, 
GSL~\cite{galassi2009}, 
and the CUBA library~\cite{Hahn:2004fe,Hahn:2014fua})
that numerically calculates dimensionally regularized integrals~\cite{Borowka:2017idc}.
pySecDec has a number of features that make it attractive for use
in QCD sum-rules analyses.
In principle, it places no restrictions on numbers of loops, external lines, or distinct masses.
Also, external momenta can take on any values in the complex plane.
pySecDec computes divergent and finite parts of an integral 
and reports uncertainties for each.
Furthermore, it can be applied to both scalar and tensor integrands.

In this article, we show that pySecDec can be successfully incorporated into 
the QCD sum-rules methodology.
We do so by considering the two-point pseudoscalar (\ie, $J^{PC}=0^{-+}$) charmonium 
hybrid correlation function. 
This correlator and its corresponding QCD Laplace sum-rules (LSRs) were computed 
in~\cite{Govaerts:1985fx,Berg:2012gd};
there, all dimensionally regularized integrals were computed analytically.
Here, instead, we use pySecDec to numerically compute the needed divergent integrals,
and then formulate the sum-rules with a straightforward numerical contour integral in the 
complex plane.
We find that the errors introduced into the LSRs from numerical integration techniques are negligible
compared to the uncertainties already present due to experimental uncertainties in the 
parameters of QCD (\eg, the strong coupling, the quark masses, and the condensates).
\add{Also, we show that numerical integration 
methods can be applied successfully to the computation of
finite-energy sum-rules (FESRs) and Gaussian sum-rules (GSRs) in addition to LSRs.}
\section{Laplace sum-rules}\label{lsrs}
To probe pseudoscalar charmonium hybrids, we use the current~\cite{Govaerts:1985fx}
\begin{equation}\label{current}
  j_{\mu} = \frac{g_s}{2} \overline{c} \gamma^{\rho}\lambda^a \tilde{G}^a_{\mu\rho}c
\end{equation}
where $c$ is a charm quark field and
\begin{equation}
  \tilde{G}^a_{\mu\rho}=\frac{1}{2}\epsilon_{\mu\rho\omega\eta}G^a_{\omega\eta}
\end{equation}
is the dual gluon field strength tensor defined using the Levi-Civita symbol
$\epsilon_{\mu\rho\omega\eta}$.
The diagonal correlator of~(\ref{current}) is defined and decomposed as follows:
\begin{align}\label{correlator}
  \Pi_{\mu\nu}(q) & = i\int\!\dfourx\, e^{iq\cdot x} \langle\Omega| \tau j_{\mu}(x) j_{\nu}^{\dagger}(0) 
    |\Omega\rangle\\
    & = \frac{q_{\mu}q_{\nu}}{q^2}\Pi_0(q^2) + \bigg(\frac{q_{\mu}q_{\nu}}{q^2}-g_{\mu\nu}\bigg)\Pi_1(q^2) \label{decomposition}
\end{align}
where $\Pi_0(q^2)$ probes spin-0 states and $\Pi_1(q^2)$ probes spin-1 states. 
We are interested in pseudoscalar hybrids and so we focus on
$\Pi_0(q^2)$ which satisfies a dispersion relation for $q^2<0$,
\begin{equation}\label{dispersion}
  \Pi_0(q^2) = q^8\int_{t_0}^{\infty}\!\frac{\frac{1}{\pi}\Im\Pi_0(t)}{t^4(t-q^2)}\,\dt + \cdots,
\end{equation}
where $t_0$ is a hadron production threshold
and $\cdots$ represents subtraction constants, a polynomial of degree-three in $q^2$.

\begin{figure}[htbp]
\begin{center}
  \includegraphics[width = 0.7\textwidth]{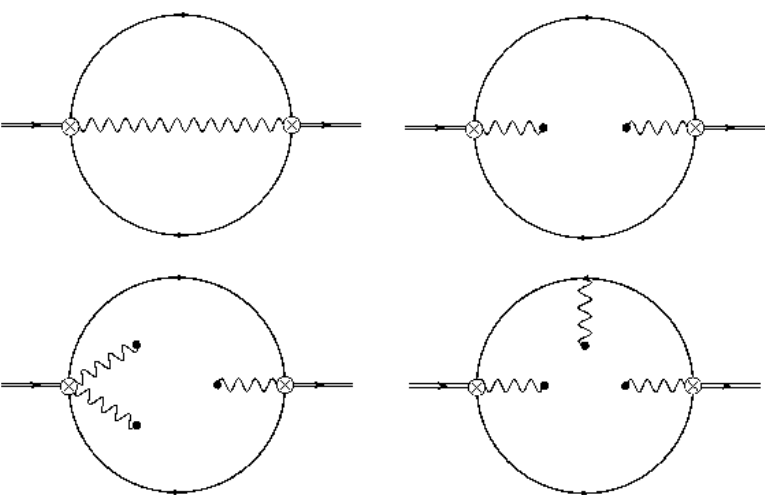}
\end{center}
\caption{\label{figureFeynmanDiagrams} The Feynman diagrams that contribute to 
  $\Pi_0^{\text{QCD}}(q^2)$ at LO in $\alpha_s$ up to and including 6d 
  condensates.  
  Feynman diagrams were drawn with JaxoDraw~\cite{BinosiTheussl2004}.}
\end{figure}

On the left-hand side of~(\ref{dispersion}), the function $\Pi_0(q^2)$ is computed using QCD
and the OPE.
In~\cite{Govaerts:1985fx,Berg:2012gd}, $\Pi_0(q^2)$ was calculated to leading-order (LO) in $\alpha_s$,
including non-perturbative terms proportional to the
four-dimensional (\ie, 4d) gluon condensate~\cite{Launer:1983ib}
\begin{equation}\label{alphaGG}
  \aGG\equiv\vev{\alpha_s G^a_{\mu\nu}G^a_{\mu\nu}} =\left(0.075\pm 0.020\right)\ \gev^4
\end{equation}
and the 6d gluon condensate~\cite{Narison2010}
\begin{equation}\label{gluonSixD}
  \gggGGG\equiv\vev{g_s^3 f^{abc} G^a_{\mu\nu} G^b_{\nu\rho} G^c_{\rho\mu}}
  =\left(\left(8.2\pm 1.0\right)\ \gev^2\right)\aGG.
\end{equation}
The corresponding diagrams are reproduced in fig.~\ref{figureFeynmanDiagrams}.
Loop integrals were handled using dimensional regularization in $D=4+2\epsilon$ dimensions
at $\overline{\text{MS}}$ renormalization scale $\mu$.
All dimensionally regularized integrals were evaluated analytically
giving
\begin{multline}\label{PiQCD}
\Pi^{\text{QCD}}_0(q^2) 
= \frac{m_c^6 \alpha_s}{270\pi^3}
\Big(
  9(4z^3-25z^2+31z-10)\, {}_3 F_2\big(1,1,1;\tfrac{3}{2},3;z\big)\\
  +z(8z^3+8z^2+29z-10)\, {}_3 F_2\big(1,1,2;\tfrac{5}{2},3;z\big)
\Big)
\\
  + \frac{m_c^2}{18\pi}z(2z+1)\, {}_2 F_1\big(1,1;\tfrac{5}{2};z\big)\langle\alpha G^2 \rangle\\
  + \frac{1}{384\pi^2(z-1)}
\Big(
  (2z^2-2z+1)\, {}_2 F_1\big(1,1;\tfrac{5}{2};z\big)
  + (10z^2-20z+7)
\Big)
\langle g^3 G^3\rangle
\end{multline}
where
\begin{equation}
  z=\frac{q^2}{4m_c^2},
\end{equation}
$m_c$ is the charm quark mass, and ${}_p F_q(\ldots;\ldots;z)$ are generalized hypergeometric functions.  
As in~\cite{Berg:2012gd}, a polynomial in $z$ has been omitted from the right-hand side of~(\ref{PiQCD})
as polynomials do not contribute to the LSRs (see~(\ref{lsrFinal})).
Note that we have added a superscript $\text{QCD}$ on the left-hand side
of~(\ref{PiQCD}) to emphasize that this quantity was computed using QCD.

%

On the right-hand side of~(\ref{dispersion}), we use a resonances-plus-continuum 
model,
\begin{equation}\label{spectral}
  \frac{1}{\pi}\Im\Pi_0(t) = \rho^{\text{had}}(t) + \theta(t-s_0)\frac{1}{\pi}\Im\Pi^{\text{QCD}}_0(t),
\end{equation}
where $\rho^{\text{had}}(t)$ represents the resonance content of $\Im\Pi_0(t)$,
$\theta(t)$ is the Heaviside step function, and $s_0$ is the continuum 
threshold parameter.

In~(\ref{dispersion}), to eliminate the (generally unknown) subtraction constants
and to accentuate the low-energy contribution of $\Im\Pi_0(t)$ to the integral,
we consider (continuum-) subtracted LSRs, $\lsr_k(\tau,\,s_0)$,
of (usually nonnegative) integer weight $k$ at Borel scale $\tau$,
\begin{equation}\label{subtracted}
  \lsr_k(\tau,\,s_0) = \lsr_k(\tau) 
  - \int_{s_0}^{\infty}\! t^k e^{-t\tau} \frac{1}{\pi}\Im\Pi_0^{\text{QCD}}(t)\,\dt,
\end{equation}
in terms of unsubtracted LSRs, $\lsr_{k}(\tau)$, defined by 
\begin{equation}\label{unsubtracted}
  \lsr_k(\tau) = \frac{1}{\tau}
  \lim_{\stackrel{N,Q^2\rightarrow\infty}{\tau=N/Q^2}}
  \frac{\big(-Q^2\big)^N}{\Gamma(N)}\bigg(\frac{d}{dQ^2}\bigg)^N
  \left\{(-Q^2)^k \Pi_0(Q^2)\right\},
\end{equation}
where $Q^2=-q^2$~\cite{Shifman:1978by,Shifman:1978bx,Reinders:1984sr}.
Equations~(\ref{dispersion}) and (\ref{spectral})--(\ref{unsubtracted}) 
together imply 
\begin{equation}\label{LSRduality}
  \lsr_k(\tau,\,s_0) = \int_{t_0}^{\infty}\! t^k e^{-t\tau} 
  \rho^{\text{had}}(t)\dt.
\end{equation}
Details on how to compute $\lsr_k(\tau,\,s_0)$ for a correlator such as~(\ref{PiQCD})
can be found in the literature
(\eg,~\cite{Shifman:1978by,Govaerts:1985fx,Berg:2012gd}).
The result can be expressed as
\begin{equation}\label{lsrFinal}
  \lsr_k(\tau,\,s_0) = \int_{4m_c^2(1+\eta)}^{s_0}\! t^k e^{-t\tau} 
    \frac{1}{\pi}\Im\Pi_0^{\text{QCD}}(t)\,\dt
  + \frac{1}{2\pi i}\int_{\Gamma_\eta}\! \left(q^2\right)^k 
  e^{-q^2\tau} \Pi_0^{\text{QCD}}(q^2)\,\dqsq
\end{equation}
where $\Gamma_\eta$ is a circle in the complex $q^2$-plane 
of arbitrary radius $\eta$ centred at $q^2=4m_c^2$ which we parameterize by
\begin{equation}
  q^2 = 4m_c^2\left( 1 + \eta e^{i\theta}\right)
\end{equation}
for $\theta_{i}=2\pi^{-}$ to $\theta_f=0^{+}$.
For $t\geq4m_c^2$, as computed in~\cite{Berg:2012gd}, we have that
\begin{multline}\label{ImOPE}
  \Im\Pi_0^{\text{QCD}}(t) = \frac{\alpha_s m_c^6}{120\pi^2 z^2}
    \Bigg(
      \sqrt{z(z-1)}(30-115z+166z^2+8z^3+16z^4)\\
      -15(-2+9z-16z^2+16z^3)\log\big(\sqrt{z-1}+\sqrt{z}\big)
    \Bigg)\\
  + \frac{m_c^2}{12}\sqrt{1-\frac{1}{z}}(2z+1)\aGG
  + \frac{1}{256\pi z(z-1)^2}\sqrt{1-\frac{1}{z}}(2z^2-2z+1)\gggGGG
\end{multline}
where
\begin{equation}
  z=\frac{t}{4m_c^2}.
\end{equation}

To implement renormalization-group improvement, we replace the strong coupling 
and the charm quark mass in~(\ref{PiQCD}) and~(\ref{ImOPE}) by 
one-loop, $\overline{\text{MS}}$ running quantities at four flavours~\cite{Narison:1981ts}:
\begin{gather}
  \alpha_{s}(\mu) = \frac{\alpha_{s}\left(m_{\tau}\right)}%
    {1+\frac{25}{12\pi}\alpha_{s}\left(m_{\tau}\right)
    \log(\frac{\mu^2}{m_{\tau}^2})} 
  \label{runningCoupling}\\
  m_c(\mu) = m_c(\overline{m}_c)\left(\frac{\alpha_s(\mu)}{\alpha_s(\overline{m}_c)}\right)^\frac{12}{25}
  \label{runningMass} 
\end{gather}
where~\cite{Olive:2016xmw} 
\begin{gather}
  m_{\tau} = (1.77686\pm 0.00012)\ \gev\\
  \alpha_s(m_{\tau}) = 0.330\pm 0.014 \label{alphatau}\\
  \overline{m}_c = (1.275\pm 0.025)\ \gev. \label{mcbar}
\end{gather}
Also, we set $\mu=\overline{m}_c$.
\section{Methods and results}\label{methods}
We compute the LSRs, $\lsr_k(\tau,\, s_0)$, (see~(\ref{lsrFinal})) of weight $k=0$ and $k=1$  
using pySecDec-determined numerical results for~$\Pi^{\text{QCD}}_0(q^2)$ 
and $\Im\Pi_0^{\text{QCD}}(t)$
rather than using the analytic expressions~(\ref{PiQCD}) and~(\ref{ImOPE}) respectively.

But first, we must decide on a value for the arbitrary parameter $\eta$ in~(\ref{lsrFinal}). 
The correlator~(\ref{PiQCD}) has a branch cut along the positive real semi-axis
originating at the branch point $z=1$, \ie, $q^2=4m_c^2$.  
Near the branch point, the 6d gluon condensate OPE term is unbounded.
Also, the imaginary part of the 4d gluon condensate term
steeply approaches the branch point from above.
Such behaviour is common and corresponds to large values of
$|\frac{\mathrm{d}}{\mathrm{d}q^2}\Pi_0^{\text{QCD}}(q^2)|$ near $q^2=4m_c^2$.
To improve the reliability of numerical integration methods,
it is advantageous to avoid these large-magnitude derivatives by
choosing a large value for $\eta$.
However, because of the exponential factor, $e^{-q^2\tau}$, in the integrand 
of the second integral on the right-hand side of~(\ref{lsrFinal}), it is also advantageous
to ensure that $\Re(q^2)> 0$ along $\Gamma_{\eta}$.
Both advantages are realized by choosing $\eta$ slightly less than 1; 
in our case, $\eta=0.999$.

Regarding $\tau$, we consider $0.1\ \gev^{-2}\leq\tau\leq 0.5\ \gev^{-2}$, a range that 
easily covers the $\tau$-interval used in the LSRs analysis of~\cite{Berg:2012gd}.
Regarding $s_0$, we restrict our attention to $s_0>8m_c^2$, consistent with our choice of $\eta$.
As for an upper bound on $s_0$, in some LSRs analyses, 
it is useful to compute $\lsr_k(\tau,\,s_0)$ as $s_0\rightarrow\infty$.  
However, because of the exponential damping factor, $e^{-t\tau}$, in the integrand of the 
first integral on the right-hand side~(\ref{lsrFinal}), 
the $s_0\rightarrow\infty$ limit can be practically calculated by simply 
evaluating $\lsr_k(\tau,\,s_0)$ at some large value of $s_0$.  
For the case under consideration, a maximum value $s_0\approx 120\ \gev^2$ 
is large enough by a comfortable margin.

\begin{figure}[htbp]
\begin{center}
  \includegraphics[width = 0.9\textwidth]{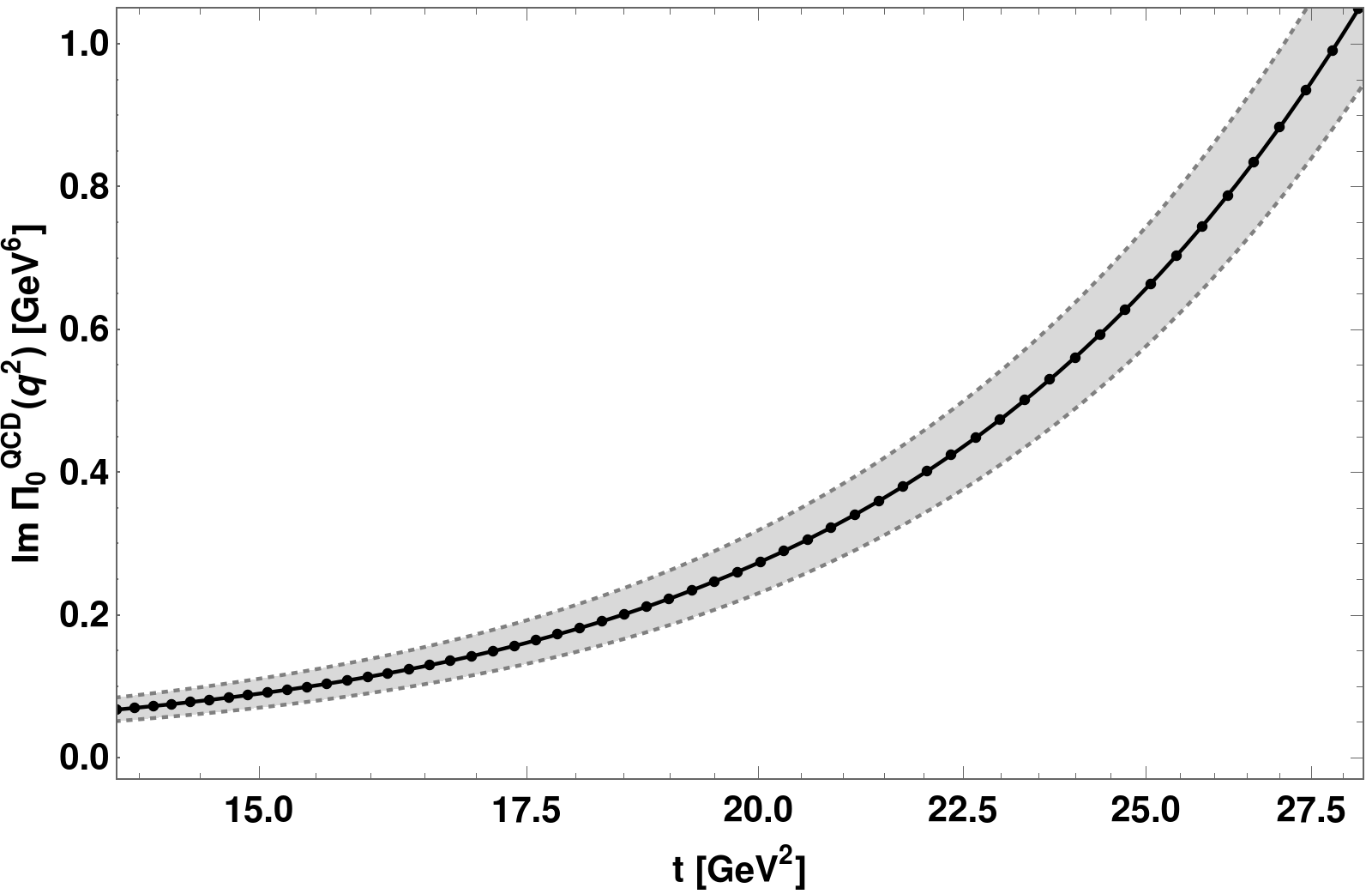}
\end{center}
\caption{\label{ImPiSmallPlot} 
  Comparison of~$\Im\Pi_0^{\text{QCD}}(t)$ calculated analytically
  (see~(\ref{ImOPE})) and numerically using pySecDec.
  (Note the logarithmic horizontal axis.)
  The solid line represents the analytic result with central values of
  $\aGG$, $\gggGGG$, $\alpha_s(m_{\tau})$, and~$\overline{m}_c$  
  (see~(\ref{alphaGG}), (\ref{gluonSixD}), (\ref{alphatau}), and~(\ref{mcbar})~respectively).
  The two dotted lines represent upper and lower bounds on the analytic result 
  that take into account uncertainties in $\aGG$, $\gggGGG$, $\alpha_s(m_{\tau})$, 
  and~$\overline{m}_c$.
  The dots represent pySecDec-generated results at those values of the grid 
  points~(\ref{tGrid}) which satisfy $t\lesssim 28\ \gev^2$.
  The relative uncertainties of the pySecDec-generated results are of the order 
  $10^{-7}$,
  and the corresponding error bars are much smaller than the dots used in the figure.
  }
\end{figure}

\begin{figure}[htbp]
\begin{center}
  \includegraphics[width = 0.9\textwidth]{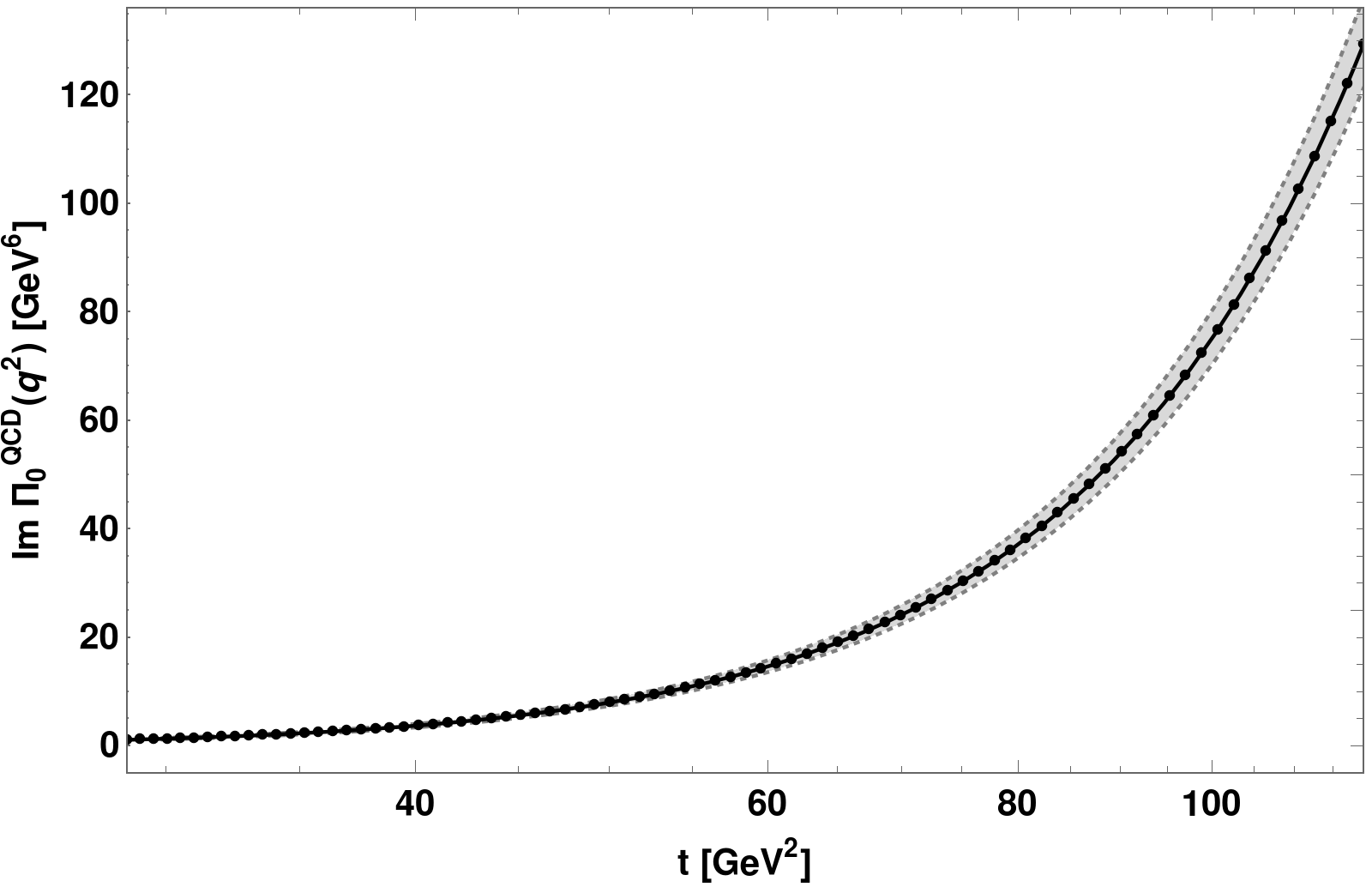}
\end{center}
\caption{\label{ImPiLargePlot} 
The same as fig.~\ref{ImPiSmallPlot} but for a larger range of $t$-values.}
\end{figure}

\begin{figure}[htbp]
\begin{center}
  \includegraphics[width = 0.9\textwidth]{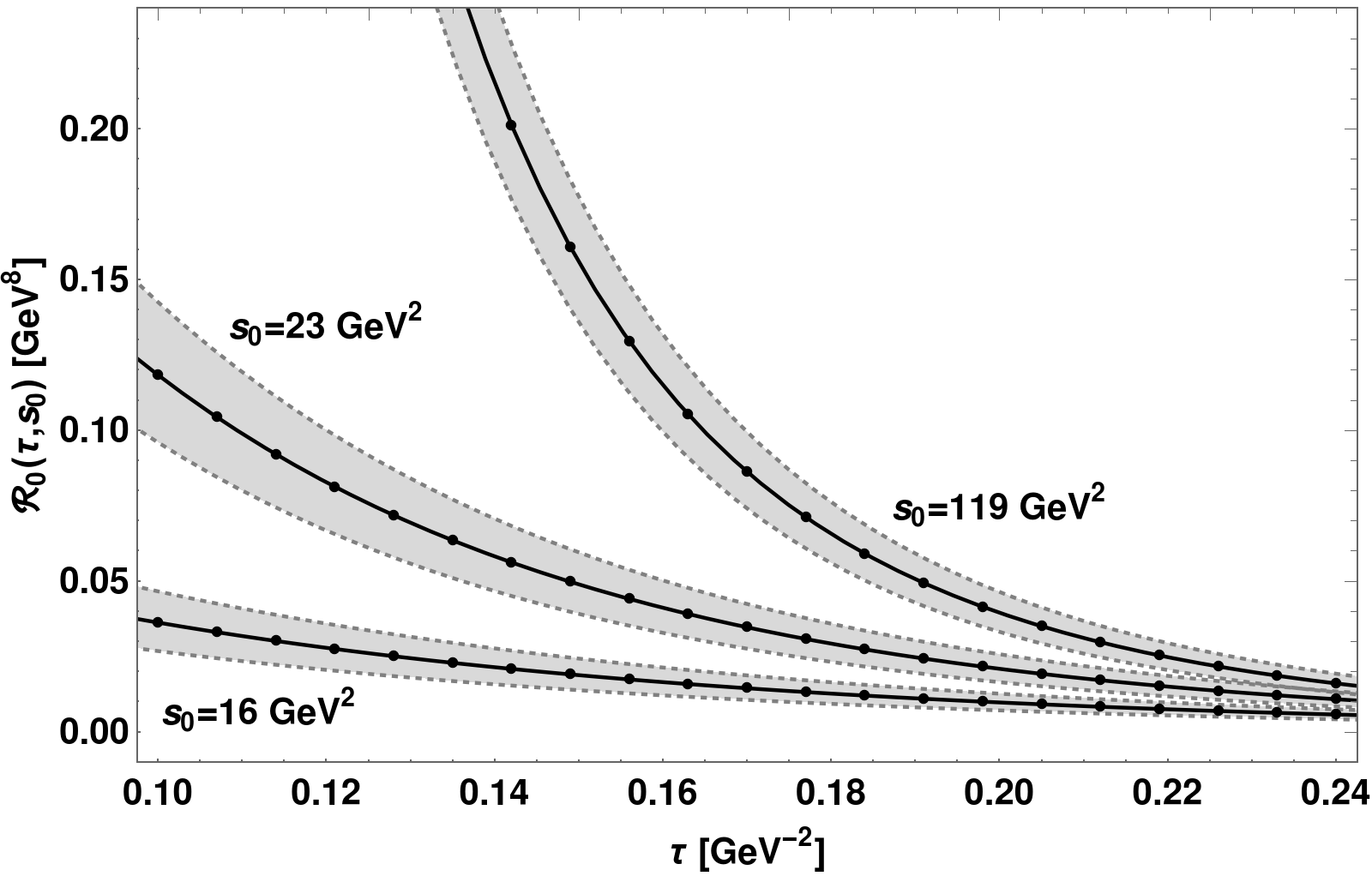}
\end{center}
\caption{\label{RzeroPlot}
  Comparison of the $k=0$ LSRs, $\lsr_0(\tau,\,s_0)$, (see~(\ref{lsrFinal}))
  calculated using analytic results for $\Pi_0^{\text{QCD}}(q^2)$ and $\Im\Pi_0^{\text{QCD}}(t)$
  (see~($\ref{PiQCD}$) and~($\ref{ImOPE}$) respectively)
  and using pySecDec-generated numerical results
  at a representative set of continuum threshold parameter values,
  $s_0=16\ \gev^2$, $s_0=23\ \gev^2$, and~$s_0=119\ \gev^2$.
  The solid lines represent LSRs computed using analytic results with central values of
  $\aGG$, $\gggGGG$, $\alpha_s(m_{\tau})$, and~$\overline{m}_c$  
  (see~(\ref{alphaGG}), (\ref{gluonSixD}), (\ref{alphatau}), and~(\ref{mcbar})~respectively).
  Dotted lines represent upper and lower bounds on the solid lines 
  that take into account uncertainties in $\aGG$, $\gggGGG$, $\alpha_s(m_{\tau})$, 
  and~$\overline{m}_c$.
  The dots represent LSRs computed at a representative discrete collection of $\tau$-values 
  using pySecDec-generated numerical results for $\Pi_0^{\text{QCD}}(q^2)$ and 
  $\Im\Pi_0^{\text{QCD}}(t)$.
  The relative uncertainties due to pySecDec are, at worst, on the order of $10^{-4}$, 
  and the corresponding error bars are much smaller than the dots in the figure.}
\end{figure}

\begin{figure}[htbp]
\begin{center}
  \includegraphics[width = 0.9\textwidth]{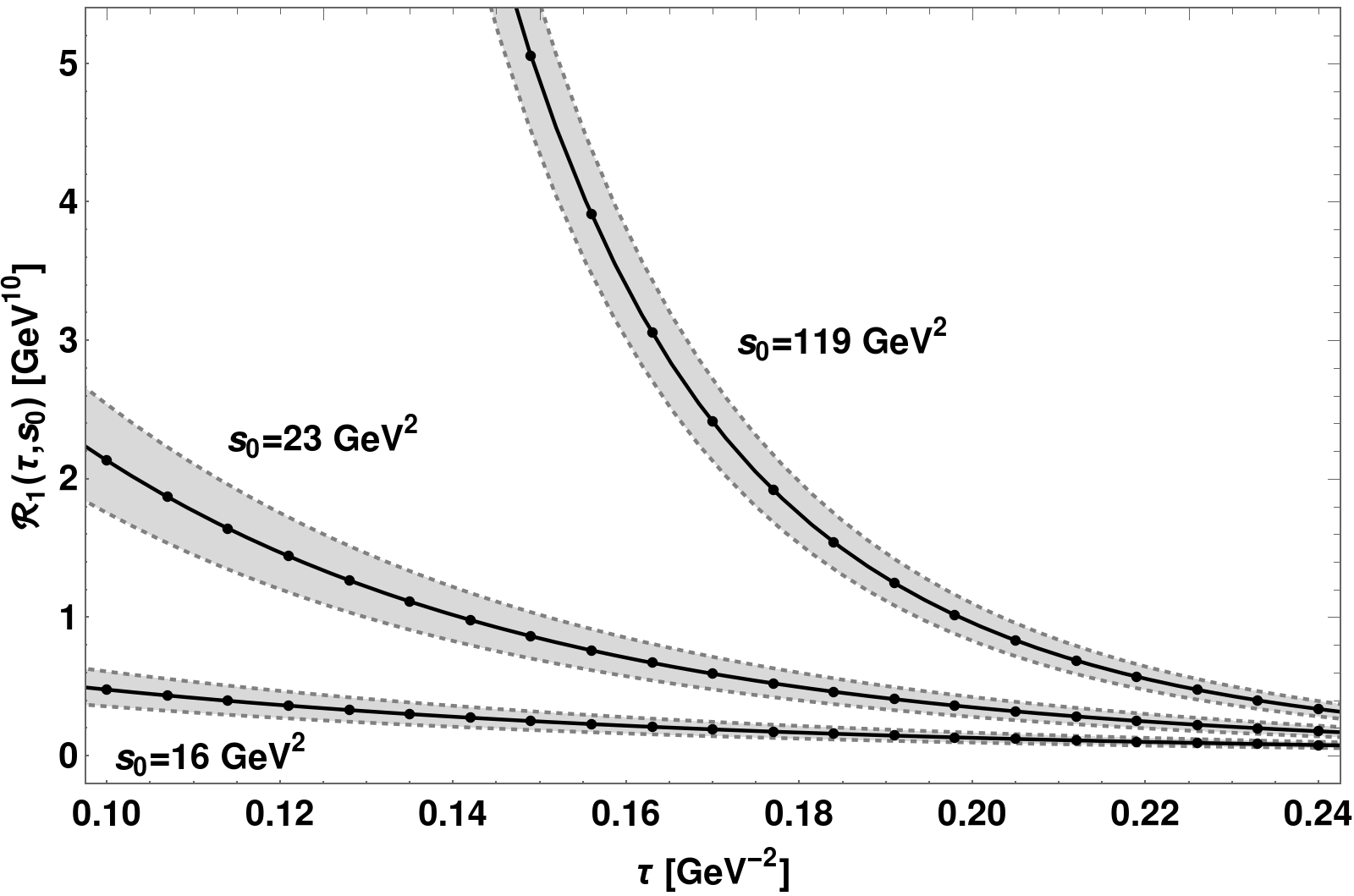}
\end{center}
\caption{\label{RonePlot}
  The same as fig.~\ref{RzeroPlot} but for $k=1$ LSRs, $\lsr_1(\tau,\,s_0)$.
}
\end{figure}

We compute each of the two integrals on the right-hand side of~(\ref{lsrFinal}) using Simpson's rule.
For the first integral, we use sample points $\{t_n\}_{n=0}^{144}$ where 
\begin{equation}\label{tGrid}
  t_n = 4m_c^2(1.02^n+\eta)
\end{equation}
which unevenly spans the range $8m_c^2\lesssim t\lesssim120\ \gev^2$. 
The adaptive step-sizing of~(\ref{tGrid}) emphasizes the behaviour of the integrand
near the lower limit of integration where the exponential damping is smallest,
and is needed for Simpson's rule to give accurate results for large $s_0$.
Note that, for $s_0<120\ \gev^2$, not all of the sample points~(\ref{tGrid}) are used.
For example, at the value $s_0=25\ \gev^2$, 
only the first 53 grid points 
are used to numerically evaluate the integral in question.
In figs.~\ref{ImPiSmallPlot} and~\ref{ImPiLargePlot}, 
we plot the analytic result for $\Im\Pi_0^{\text{QCD}}(t)$ from~(\ref{ImOPE}) 
along with corresponding numerical pySecDec-calculated results at 
the sample points~(\ref{tGrid}).
For the second integral on the right-hand side of~(\ref{lsrFinal}), we use 
123 sample points evenly spaced along the circle $\Gamma_{\eta}$.
Because
\begin{equation}
  \Pi_0^{\text{QCD}}\big((q^{2})^{*}\big) = \Big(\Pi_0^{\text{QCD}}(q^2)\Big)^{*},
\end{equation}
we only need to use pySecDec to evaluate $\Pi_0^{\text{QCD}}(q^2)$ for 
those sample points $q^2$ satisfying $\Im(q^2)\geq0$.
We don't directly compare the analytic result for $\Pi_0^{\text{QCD}}(q^2)$ in~(\ref{PiQCD}) 
to pySecDec-calculated results because (as noted in sect.~\ref{lsrs})
an irrelevant polynomial has been omitted from~(\ref{PiQCD}).
In fig.~\ref{RzeroPlot}, we compare the $k=0$ LSRs~(\ref{lsrFinal})
calculated using analytic results for $\Pi_0^{\text{QCD}}(q^2)$ and $\Im\Pi_0^{\text{QCD}}(t)$ 
(\ie,~(\ref{PiQCD}) and~(\ref{ImOPE}) respectively) 
and using pySecDec-generated results at a selection of $s_0$-values.
Figure~\ref{RonePlot} is analogous to fig.~\ref{RzeroPlot}, but for $k=1$ rather 
than $k=0$ LSRs.


\section{Discussion}

From figs.~\ref{ImPiSmallPlot} and~\ref{ImPiLargePlot}, we see that there is exceptional 
agreement between $\Im\Pi_0^{\text{QCD}}(t)$ computed analytically (see~(\ref{ImOPE}))
and numerically using pySecDec over the entire range $8m_c^2\lesssim t \lesssim 120\ \gev^2$.
The relative errors between true values (\ie, those computed analytically)
and pySecDec-calculated values are of the order $10^{-9}$.
Also, for the pySecDec-calculated results (\ie, the dots in the two figures), the
relative uncertainties as reported by pySecDec are of the order $10^{-7}$,
and the corresponding error bars are far too small to be 
seen in the two figures.
And so, the pySecDec-generated errors are much smaller than the pySecDec-generated
uncertainties which, in turn, are negligible
compared to the uncertainties in $\Im\Pi_0^{\text{QCD}}(t)$
associated with experimental uncertainties in the QCD parameters, 
$\aGG$, $\gggGGG$, $\alpha_s(m_{\tau})$, and~$\overline{m}_c$ 
(see~(\ref{alphaGG}), (\ref{gluonSixD}), (\ref{alphatau}), and~(\ref{mcbar})~respectively).
In principle, the uncertainties attributable to pySecDec could
be reduced at the cost of longer pySecDec runtimes. 
Note that the total pySecDec runtime needed to produce the data points in 
figs.~\ref{ImPiSmallPlot} and~\ref{ImPiLargePlot} was a few hours 
on a mid-range laptop.

Similarly, from figs.~\ref{RzeroPlot} and~\ref{RonePlot}, we see that there is excellent 
agreement between the LSRs, $\lsr_0(\tau,\,s_0)$ and $\lsr_1(\tau,\,s_0)$, computed
using the analytic results~(\ref{PiQCD}) and~(\ref{ImOPE}) and pySecDec-calculated 
equivalents.
The relative errors between true values and values determined numerically are of the order $10^{-5}$ or less.
This conclusion holds for the entire range of $s_0$-values considered 
(\ie, $8m_c^2\lesssim s_0 \lesssim 120\ \gev^2$) and the entire range of
$\tau$-values considered (\ie, $0.1\ \gev^{-2}\leq\tau\leq 0.5\ \gev^{-2}$),
not just at the particular $s_0$-values and $\tau$-values used in
figs.~\ref{RzeroPlot} and~\ref{RonePlot}.
The relative uncertainties in the numerically-computed
LSRs attributable to pySecDec are of the order $10^{-4}$ or less.
Of course, there are also uncertainties in the numerically-computed LSRs
associated with Simpson's rule.  We have made no attempt to quantify 
these uncertainties, but note that they decrease rapidly with increasing
numbers of sample points.
And so, 
computing here the LSRs $\lsr_0(\tau,\,s_0)$ and $\lsr_1(\tau,\,s_0)$
using a combination of pySecDec and Simpson's rule 
produced errors that were an order of magnitude smaller than the 
uncertainties attributable to pySecDec which, in turn, were negligible
compared to the uncertainties attributable to the parameters of QCD,
$\aGG$, $\gggGGG$, $\alpha_s(m_{\tau})$, and~$\overline{m}_c$.

\add{
In addition to LSRs, there are other variants of QCD sum-rules
used in the literature.  Two such variants are FESRs,
$\fesr_k(s_0)$,~\cite{Shankar:1977ap,Narison:2007spatmp}
\begin{equation}\label{fesr}
  \fesr_k(s_0) = \int_{4m_c^2(1+\eta)}^{s_0}\! t^k 
    \frac{1}{\pi}\Im\Pi_0^{\text{QCD}}(t)\,\dt
  + \frac{1}{2\pi i}\int_{\Gamma_\eta}\! \left(q^2\right)^k 
  \Pi_0^{\text{QCD}}(q^2)\,\dqsq
\end{equation}
and GSRs, $\gsr_k(\hat{s},\,\tau,\,s_0)$,~\cite{Bertlmann:1984ih,Narison:2007spatmp,Ho:2018}
\begin{equation}\label{gsr}
  \gsr_k(\hat{s},\,\tau,\,s_0) = \int_{4m_c^2(1+\eta)}^{s_0}\! t^k 
  \frac{e^{-\frac{(\hat{s}-t)^2}{4\tau}}}{\sqrt{4\pi\tau}} 
    \frac{1}{\pi}\Im\Pi_0^{\text{QCD}}(t)\,\dt
  + \frac{1}{2\pi i}\int_{\Gamma_\eta}\! \left(q^2\right)^k 
  \frac{e^{-\frac{-(\hat{s}-q^2)^2}{4\tau}}}{\sqrt{4\pi\tau}}
  \Pi_0^{\text{QCD}}(q^2)\,\dqsq.
\end{equation}
The difference between LSRs~(\ref{lsrFinal}) and either FESRs~(\ref{fesr}) 
or GSRs~($\ref{gsr}$) is the kernel of the integrands, 
\ie, a decaying exponential in LSRs, a trivial kernel in FESRs, and 
a Gaussian in GSRs.
On the left-hand side of~(\ref{gsr}), in addition to the continuum threshold $s_0$, 
there are two other independent variables: $\hat{s}$, the position of the 
peak of the Gaussian kernel (in $\gev^2$)
and $\tau$, the variance of the Gaussian kernel (in $\gev^4$).
(Despite the same letter $\tau$ being used in the definitions
of LSRs and GSRs, the two parameters are not the same.)
The pySecDec-determined correlator results for $\Pi_0^{\text{QCD}}(q^2)$
and $\Im\Pi_0^{\text{QCD}}(t)$
can be used to numerically evaluate~(\ref{fesr}) and~(\ref{gsr}) using the same
procedure as that described in Section~\ref{methods} for LSRs.
The integration contour of Section~\ref{methods} was chosen with LSRs in mind;  
nevertheless, we find excellent agreement between FESRs and GSRs computed using
analytic correlator results and pySecDec-computed correlator results.
For FESRs, we find relative errors less than $10^{-5}$ for $k\in\{0,\,1\}$ over 
the entire range of $s_0$ values considered, \ie\ $8m_c^2\lesssim s_0\lesssim 120~\gev^2$.
For GSRs, we find relative errors less than $10^{-4}$ for $k\in\{0,\,1\}$ over 
the entire range of $s_0$ values considered and where 
$10\ \gev^4\leq\tau\leq 20\ \gev^4$ (a range consistent with the discussion
in~\cite{Ho:2018}) and
$-5\ \gev^2\leq\hat{s}\leq 40\ \gev^2$ (a range that covers essentially all of
the area underneath $\gsr_k(\hat{s},\,\tau,\,s_0)$ for the range of $\tau$ values 
considered and for realistic values of $s_0$.)
}

For the pseudoscalar charmonium hybrid correlator
considered in this article~($\Pi_0$ on the right-hand side of~(\ref{decomposition})), 
the dimensionally-regularized Feynman diagrams of fig.~\ref{figureFeynmanDiagrams} 
could be evaluated analytically yielding~(\ref{PiQCD}).
Corresponding LSRs\add{, FESRs, and GSRs} 
could then be computed in a straightforward fashion.
In this article, as an alternative, we explicitly showed that
\add{QCD sum-rules}
computed numerically using a combination of pySecDec and Simpson's rule
were in excellent
agreement with those constructed from an analytic expression for $\Pi_0$.
As Feynman diagrams get increasingly complicated due to larger numbers
of loops, external lines, and/or masses, evaluating them analytically can 
become prohibitively difficult. 
In such cases, evaluating the diagrams and
computing corresponding \add{QCD sum-rules} with the help of pySecDec is a viable option. 

\clearpage
\section*{Acknowledgements}
We are grateful for financial support from the National Sciences and 
Engineering Research Council of Canada (NSERC).
Also, we would like to thank T.~G.~Steele for many helpful discussions concerning this project.

\bibliographystyle{h-physrev}
\bibliography{research}

\end{document}